\documentclass[twocolumn,aps,pra,showpacs,floatfix,superscriptaddress,amssymb,10pt]{revtex4-1}

\usepackage{graphicx}
\usepackage{dcolumn}
\usepackage{bm}
\usepackage{booktabs}
\usepackage{adjustbox}
\usepackage{graphicx} 
\usepackage{amsmath}   
\usepackage{color}
\usepackage{float}
\usepackage{array}
\usepackage{graphicx}
\usepackage[caption = false]{subfig}

\begin{document}
\preprint{}
\title{High precision calculations of the $ 1s^{2} 2s 2p$  $^1P_{1} \,-\ 1s^{2} 2s^{2}$ $^1S_{0} $ 
       spin allowed $E1$ transition in C III}

\author{M.~Bilal}
\affiliation{Helmholtz-Institut Jena, Fr\"o{}belstieg 3, D-07743 Jena, Germany}
\affiliation{GSI Helmholtzzentrum f\"ur Schwerionenforschung GmbH, Planckstrasse 1, D-64291 Darmstadt, Germany}
\affiliation{Theoretisch-Physikalisches Institut, Friedrich-Schiller-Universit\"at Jena, Max-Wien-Platz 1, D-07743
Jena, Germany}

\author{A.~V.~Volotka}
\affiliation{Helmholtz-Institut Jena, Fr\"o{}belstieg 3, D-07743 Jena, Germany}
\affiliation{GSI Helmholtzzentrum f\"ur Schwerionenforschung GmbH, Planckstrasse 1, D-64291 Darmstadt, Germany}

\author{R. Beerwerth}
\affiliation{Helmholtz-Institut Jena, Fr\"o{}belstieg 3, D-07743 Jena, Germany}
\affiliation{GSI Helmholtzzentrum f\"ur Schwerionenforschung GmbH, Planckstrasse 1, D-64291 Darmstadt, Germany}
\affiliation{Theoretisch-Physikalisches Institut, Friedrich-Schiller-Universit\"at Jena, Max-Wien-Platz 1, D-07743
Jena, Germany}

\author{J. Rothhardt}
\affiliation{Helmholtz-Institut Jena, Fr\"o{}belstieg 3, D-07743 Jena, Germany}
\affiliation{GSI Helmholtzzentrum f\"ur Schwerionenforschung GmbH, Planckstrasse 1, D-64291 Darmstadt, Germany}
\affiliation{Fraunhofer Institute for Applied Optics and Precision Engineering, Albert-Einstein-Stra\ss e 7, 07745 Jena, Germany}

\author{V. Hilbert}
\affiliation{Helmholtz-Institut Jena, Fr\"o{}belstieg 3, D-07743 Jena, Germany}
\affiliation{GSI Helmholtzzentrum f\"ur Schwerionenforschung GmbH, Planckstrasse 1, D-64291 Darmstadt, Germany}
\affiliation{Fraunhofer Institute for Applied Optics and Precision Engineering, Albert-Einstein-Stra\ss e 7, 07745 Jena, Germany}

\author{S.~Fritzsche}
\affiliation{Helmholtz-Institut Jena, Fr\"o{}belstieg 3, D-07743 Jena, Germany}
\affiliation{GSI Helmholtzzentrum f\"ur Schwerionenforschung GmbH, Planckstrasse 1, D-64291 Darmstadt, Germany}
\affiliation{Theoretisch-Physikalisches Institut, Friedrich-Schiller-Universit\"at Jena, Max-Wien-Platz 1, D-07743
Jena, Germany}

\begin{abstract}

  Large-scale relativistic calculations are performed for the transition energy and line strength of the 
  $ 1s^{2} 2s 2p$  $^1P_{1} \,-\ 1s^{2} 2s^{2}$ $^1S_{0} $ transition in Be-like carbon.
  Based on the multiconfiguration Dirac-Hartree-Fock~(MCDHF) approach, different correlation models are 
  developed to account for all major electron-electron correlation contributions.
  These correlation models are tested with various sets of the initial and the final state wave functions.
  The uncertainty of the predicted line strength due to missing correlation effects is estimated from the
  differences between the results obtained with those models.
  The finite nuclear mass effect is accurately calculated taking into account the energy, wave functions
  as well as operator contributions.
  As a result, a reliable theoretical benchmark of the $E1$ line strength is provided to support high precision
  lifetime measurement of the $ 1s^{2} 2s 2p$  $^1P_{1} $ state in Be-like carbon.

\end{abstract}  
\maketitle

\section{Introduction}\label{intr}

Our understanding of the structure and dynamics of many-electron atoms and ions depends on a detailed
analysis and comparison of theoretical predictions with experimental observations of atomic properties.
Two important and complementary properties of atomic states are transition energies and transition rates.
For transition energies, the present experimental accuracy reaches the order of $10^{-6} - 10^{-18}$
\cite{Zhang.3.189.2016,Crespo.723.012052.2016,Bermuth.6.040059.2018}. For this case the interplay between
experiment and theory has improved drastically our understanding of different effects, e.g.,
the Breit interaction, finite nuclear mass, and quantum electrodynamics~(QED) effects 
\cite{Volotka.525.636.2013, Shabaev.239.60.2018}. This interplay also has great potential in the search 
for new physics \cite{Safronova.90.025008.2018}.

\begin{figure*}
\includegraphics[scale=1]{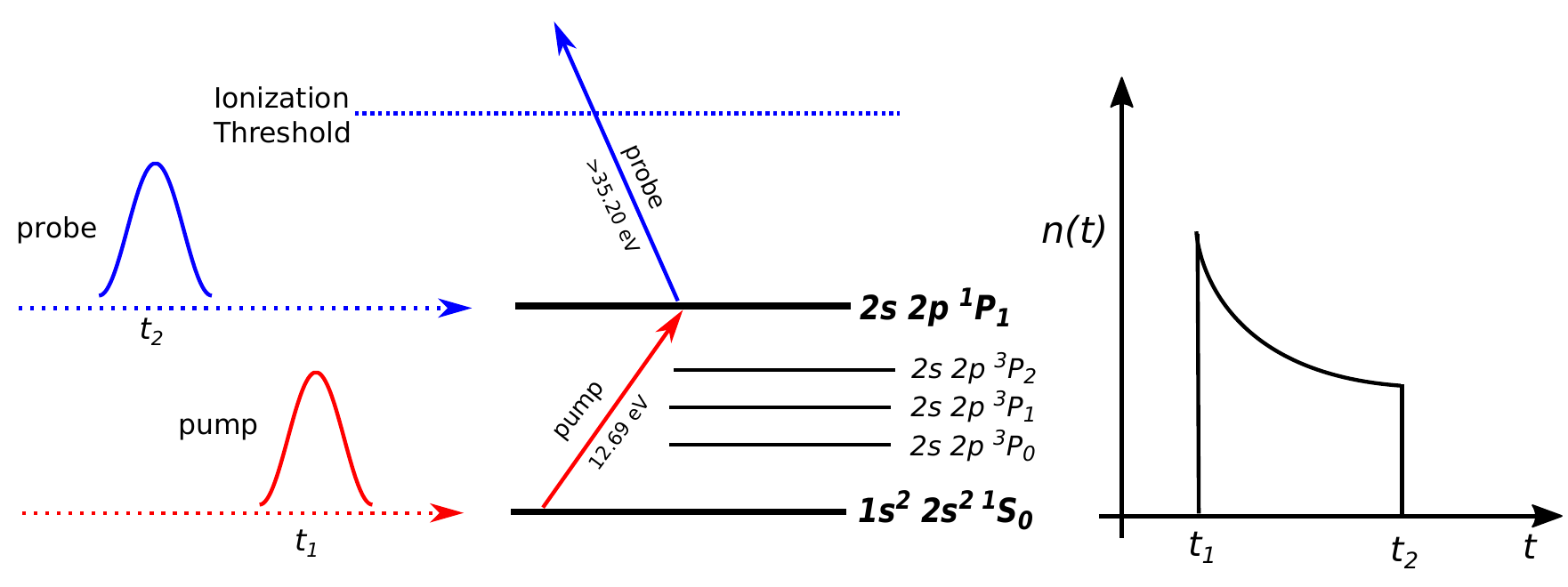}
\caption{Scheme of a pump-probe experiment for atomic lifetime measurements of the $1s^{2}2s2p$ $^1P_{1}$
state in Be-like carbon. At time $t_1$ the XUV-pump pulse excites the $1s^{2}2s^{2}$ $^1S_{0}$ ground state
of the Be-like carbon sample to the $ 1s^{2}2s2p$ $^1P_{1}$ excited state. The population $n(t)$ of the excited
state $^1P_1$ decays then exponentially. A second and temporally delayed XUV-probe pulse probes the
rest population of the excited state at time $t_2$ by ionizing Be-like carbon ion and maps out the exponential 
curve from which the decay rate is determined.}
\label{pump-probe}
\end{figure*}

For transition rates of many-electron atoms and ions, in contrast, most if not 
all the experiments provide uncertainties in the region of $30\% - 1\%$, e.g., see the reviews 
\cite{Trabert.2.15.2014,Trabert.43.074034.2010} and references therein. Whereas only a few 
experiments provide the uncertainty in the region $1\% - 0.1\%$ with some rare and favorable 
circumstances mainly for the $M1$ forbidden transitions, e.g., Refs.~\cite{Lapierre.95.183001.2005,
Brenner.75.032504.2007}. 
In this context, high hopes are pinned on the femtosecond laser technology \cite{Zewail.104.5660.2000},
which has already demonstrated great success in studies of
chemical reactions, wave function dynamics, photoionization time delays, etc. The femtosecond 
laser technology allows to perform the highly accurate pump-probe atomic lifetime measurements. In
particular, the so-called pump-probe technique has been used already in the lifetime measurements 
of the $6P_{3/2}$ excited-state in cesium atom which is relevant to the atomic parity non-conservation
\cite{Sell.84.010501.2011,Patterson.91.012506.2015}. In contrast to
neutral atoms, transitions to excited states in ions quickly reach the XUV- or X-ray energy range and, therefore,
for the pumping and/or probing processes a high-photon flux of XUV- or X-ray sources is required.
For this purpose, for instance, the Linac Coherent Light Source has been employed in the measurement
of lifetimes in Ne-like iron  \cite{Trabert.114.167.2014}. Recently, it has been also
proposed to use a compact high-power XUV-ray source in a combination with the storage ring at
GSI to perform precision spectroscopy and lifetime measurements of ions \cite{Rothhardt.2019}. 
For this purpose, a novel high-photon flux XUV-radiation source based
on the high harmonic generation in argon has been developed, which provides $\sim$100 femtosecond pulses 
at photon energies up to 26.6 eV \cite{Demmler.38.5051.2013, Rothhardt.112.233002.2014}.
As the first experiment, the measurement of the lifetime of the $1s^{2}2s2p$ $^1P_{1}$ state 
in Be-like carbon is proposed. The schematic diagram of the proposed experiment is shown in 
Fig.~\ref{pump-probe}. In principle, the relative accuracy could reach the 
order $10^{-4}-10^{-5}$ \cite{Rothhardt.2019}.

The excited state $ 1s^{2} 2s 2p$  $^1P_{1}$ decays to the $1s^{2} 2s^{2}$ $^1S_{0} $ ground state through a 
strong spin allowed $E1$ transition. Therefore, the lifetime of the $1s^2 2s 2p$  $^1P_1$ state is defined by 
the line strength of this strong transition. During past years, various calculations have been reported for this 
line strength. Among these \textit{ab initio} theories, particularly for the last three decades, are 
multiconfiguration Hartree-Fock (MCHF)~\cite{Fischer.49.323.1994}, multiconfiguration Dirac-Hartree-Fock (MCDHF) 
\cite{Ynnerman.51.2020.1995, Jonsson.57.4967.1998,Wang.234.40.2018}, many-body perturbation theory (MBPT) 
\cite{Safronova.59.286.1999}, configuration-interaction~(CI) method based on B-spline basis 
\cite{Chen.64.042507.2001} and configuration-interaction
and many-body perturbation theory~(CI+MBPT) \cite{Savukov.70.042502.2004}. As a result, the most accurate theoretical
calculations \cite{Jonsson.57.4967.1998,Chen.64.042507.2001} report an accuracy of the order $5 \times 10^{-4}$.
In view that the expected experimental accuracy is much better, there is a need for further improvements in the
theoretical calculations.

Here, we present a detailed calculation of the line strength of the $1s^{2}2s2p$  $^1P_{1} \,-\ 1s^{2}
2s^{2}$ $^1S_{0}$ transition in Be-like carbon on account of high precision experiment. We develop
various electron correlation models and use orthogonal and nonorthogonal sets of orbitals for the initial
and final states in these correlation models. It is found, that the accuracy assessment based on an
agreement between the gauges might significantly lead to underestimate the uncertainty. For this reason, we
estimate the uncertainty from the differences between the results obtained within all the correlation
models developed. In addition, the finite nuclear mass effect on the line strength is evaluated and its
gauge invariance is demonstrated after taking into account the recoil correction to the transition operator.
As a result, the calculated line strength amounts to $2.43926(37)$ with the relative accuracy $1.5 \times
10^{-4}$.

The following parts of the paper are structured as follows. In Sec. \ref{MCDHF}, we present the underlying 
theory for the calculation of the transition energy and line strength. Details of the correlation models and 
results obtained are explained in Sec. \ref{models}. In Sec. \ref{rec}, we present theoretical
methods for the finite nuclear mass effect. In the final section, we compare the obtained results with 
other theories and experiments and present the conclusion.

The atomic units $(\hbar = 1,\,e = 1,\,m_e = 1)$ are used throughout the paper unless stated otherwise.

\section{theoretical methods}\label{MCDHF}

To effectively evaluate the electron-electron correlation effects, we apply systematically enlarged
many-electron wave functions by using the general purpose relativistic atomic structure package
\textsc{Grasp2K} \cite{Jonsson.184.2197.2013}. This package implements the multiconfiguration
Dirac-Hartree-Fock~(MCDHF) method in $jj$-coupling \cite{Grant.b.2007,Fischer.49.182004.2016}. In this
method, the wave function $\Psi$ of a state labelled $\Gamma $, total angular momentum quantum number $ J $
and parity $ \pi $ is referred to as an atomic state function~(ASF) which is represented as
$\Psi(\Gamma; \pi J)$. It is an approximate eigenfunction of the Dirac-Coulomb Hamiltonian given by
\begin{equation}
\hat{H}_{\rm DC}=\sum_{i=1}^{N}\bigg[c\bm {\alpha}_{i}\cdot \bm {p}_{i}+(\beta_{i}-1)c^{2}-V(r_{i})\bigg]+
\sum_{i<j}^{N}\frac{1}{r_{ij}}\,,
\label{eq:H_DC}
\end{equation}
where $ c $ is the speed of light, $\bm {\alpha} $ and $\beta$ are (4$\times$4) Dirac-matrices, $ V(r_{i}) $
is the potential of a two parameter Fermi nuclear charge distribution and $r_{ij}$ is the distance between
electrons $i$ and $j$.

The ASF $\Psi(\Gamma; \pi J)$ is expanded in the basis of configuration state functions~(CSFs)
of the same symmetry:
\begin{eqnarray}
\Psi{(\Gamma ; \pi J)}  = \displaystyle \sum_{j=1}^{n_{c}}c_{j}\Phi(\gamma_{j}; \pi J),
\end{eqnarray}
where $n_{c}$ is the number of CSFs, $ c_{j} $ are the mixing coefficients and $ \gamma_{j} $ denotes the
orbital occupancy and angular coupling scheme of the $j$-th CSF. The CSFs $\Phi(\gamma_{j}; \pi J)$ are a 
linear combination of Slater determinants of one electron Dirac orbitals, 
\begin{eqnarray}
\phi_{n\kappa,m}(\bm r) = \frac{1}{r}
\begin{pmatrix}
P_{n \kappa}\left(r\right)\chi_{\kappa}^{m}(\theta,\varphi) \\
iQ_{n \kappa}\left(r\right)\chi_{-\kappa}^{m}(\theta,\varphi)
\end{pmatrix}.\label{eq:2}
\end{eqnarray}
Here, $ \kappa $ is the relativistic angular momentum quantum number, $P_{n\kappa}\left(r\right)$ and
$Q_{n\kappa}\left(r\right)$ are 
the large and small radial components of the one electron wave functions represented on a logarithmic 
grid, and $\chi_{\kappa}^{m}$ is the spinor spherical harmonic. 
The radial part of the Dirac orbitals and the expansion coefficients $ c_{j} $ are optimized to self 
consistency from a set of equations which result from applying the variational principle in Dirac-Coulomb
approximation \cite{Dyall.55.425.1989}. Here we have a choice of simultaneous or separate optimization of the
orbitals for the desired ASFs. In the optimal level~(OL) scheme a variational functional is constructed to
minimize the energy for only one ASF, whereas in the extended optimal level~(EOL) scheme the calculations
can be extended to include several ASFs. In the latter case, the energy functional contains weights for 
the levels under consideration. 

The line strength of the transition is defined as a square of the reduced nondiagonal matrix element of
the electromagnetic operator:
\begin{eqnarray}
S = \left|\left\langle\Psi(\Gamma; \pi J)||\boldsymbol{T}||\Psi(\Gamma'; \pi' J')\right\rangle\right|^2\,,
\end{eqnarray}
which after the optimization of the wave functions for the states $\Psi(\Gamma; \pi J)$ and
$\Psi(\Gamma'; \pi' J')$ is calculated as
\begin{eqnarray}
S = \Bigl| \sum_{j,k} c_{j} c'_{k}\left\langle\Phi(\gamma; \pi J)||\boldsymbol{T}||\Phi(\gamma'; \pi' J')\right\rangle
    \Bigr|^2\,.
\label{ampl}
\end{eqnarray}
Here, $\boldsymbol{T}$ is $E1$ transition operator \cite{Grant.b.2007} 
\begin{eqnarray}
T^l_M &=& \frac{\sqrt{2\pi}c}{\omega}\sum_i^N\Bigl[-\sqrt{6}j_1(\omega r_i/c)Y_{1M}(\bm{n}_i)\nonumber\\
      &+&  3j_2(\omega r_i/c)c\bm{\alpha}_i\cdot\bm{Y}^2_{1M}(\bm{n}_i)\Bigr]
\label{T_l}
\end{eqnarray}
in the length gauge and
\begin{eqnarray}
T^v_M &=& \frac{\sqrt{2\pi}c}{\omega}\sum_i^N \Bigl[-\sqrt{2}j_0(\omega r_i/c)c\bm{\alpha}_i\cdot\bm{Y}^0_{1M}(\bm{n}_i)\nonumber\\
      &+&  j_2(\omega r_i/c)c\bm{\alpha}_i\cdot\bm{Y}^2_{1M}(\bm{n}_i)\Bigr]
\label{T_v}
\end{eqnarray}
%
in the velocity gauge. Here, $\omega$ is the transition energy, $Y_{JM}$ are the spherical harmonics,
$\bm{Y}^L_{JM}$ are the spherical vectors, and $j_J(\omega r/c)$ is the spherical Bessel function. 
The operators~\eqref{T_l} and \eqref{T_v} are used for the present calculations of the line strengths in length and 
velocity gauges. However, in order to investigate the dependence of the line strength on the transition energy we expand the
Bessel functions $j_J(\omega r/c) \approx (\omega r/c)^J/(2J+1)!!$, the so-called long-wavelength
approximation $\omega/c \ll 1$, and retain only the leading term in the power series expansion.
In such a way the line strength takes a form:
\begin{eqnarray}\label{eq:sl}
S^l \approx \left|\left\langle\Psi(\Gamma; \pi J)||\sum_i^N \bm{r}_i||\Psi(\Gamma'; \pi' J')\right\rangle\right|^2
\end{eqnarray}
in the length gauge and
\begin{eqnarray}\label{eq:sv}
S^v \approx \frac{c^4}{\omega^2} \left|\left\langle\Psi(\Gamma; \pi J)||\sum_i^N
    \bm{\alpha}_i||\Psi(\Gamma'; \pi' J')\right\rangle\right|^2 
\end{eqnarray}
%
in the velocity gauge. From the expressions~ (\ref{eq:sl}) and (\ref{eq:sv}) one can see, that the 
leading term of the line strength in the length form is insensitive to the transition energy whereas in the velocity form it is proportional to $\omega^{-2}$. Based on these observations one can introduce the semi-empirical
correction to the line strength in the velocity gauge by adjusting the transition energy to a more
accurate, e.g., experimental, value, i.e., $\Delta S^v = (\omega^2 - \omega_{\rm exp}^2)/
\omega_{\rm exp}^2\,S^v$. Such a correction allows to take partially into account the missing correlation
contributions. The line strength in the velocity gauge adjusted in this way $S^v_{({\rm exp})}$ is
typically much closer to the value in the length gauge $S^l$. Generally, the gauge invariance should
be restored when all correlation effects are taken into account, both for the transition matrix element
and transition energy, which was explicitly demonstrated in the framework of the relativistic many-body
perturbation theory \cite{Savukov.62.052506.2000} and QED formalism \cite{Indelicato.69.062506.2004}. 
In view of this, the excellent agreement between the gauges after adjustment suggests that the 
remaining unaccounted correlation effects to the transition amplitude are rather small. As a result, 
the difference between the line strengths calculated in the length gauge and adjusted value in 
the velocity gauge is employed for the theoretical error 
estimation \cite{Fischer.T134.014019.2009,Fischer.43.074020.2010}. However, it is still possible
that the remaining unaccounted correlation effects not only reduce the discrepancy between the gauges but also shift both
values by an amount, which is much larger than the difference between the gauges after adjustment.

\section{correlation models}\label{models}

In a view of an absence of strong criteria for the uncertainty estimation of the calculated line
strength, we performed the MCDHF calculations for different correlations models. Among
those models, we choose only four models based on the accuracy criterion for the transition energy
as it is compared with the experimental energy. These four models were based on separate and
simultaneous (orthogonal and nonorthogonal) set of orbitals for the ground and excited states. 
In each model, the correlations were incorporated by systematically extending the 
calculations in a series of steps. As a first step, the calculations were performed for 
the lowest order of approximation where the orbitals belonging to so-called reference 
configurations were spectroscopically optimized, i.e., orbitals were required to have a node 
structure similar to the corresponding hydrogenic orbitals \cite{Fischer.49.182004.2016}.
Here the reference configurations for the first three models were \{$1s^2 2s^2$, $1s^2 2p^2$\} 
for the $^1S_0$ ground state and \{$1s^2 2s2p$\} for the $^1P_1$ excited state. For the 
fourth model, the reference configurations were increased and we explain its details later in this
section. In the latter steps, the calculations were extended by expanding the basis set of CSFs using 
the active set approach. In this approach, the correlations are incorporated by virtually
exciting the electrons from spectroscopic reference configurations to a set of orbitals 
called the active set of orbitals. We increased the active set by adding a layer of correlation 
orbitals but optimized only the outermost layer and kept the remaining orbitals fixed from the
previous step of calculations.

\begin{figure}[bh]
\includegraphics[scale=0.52]{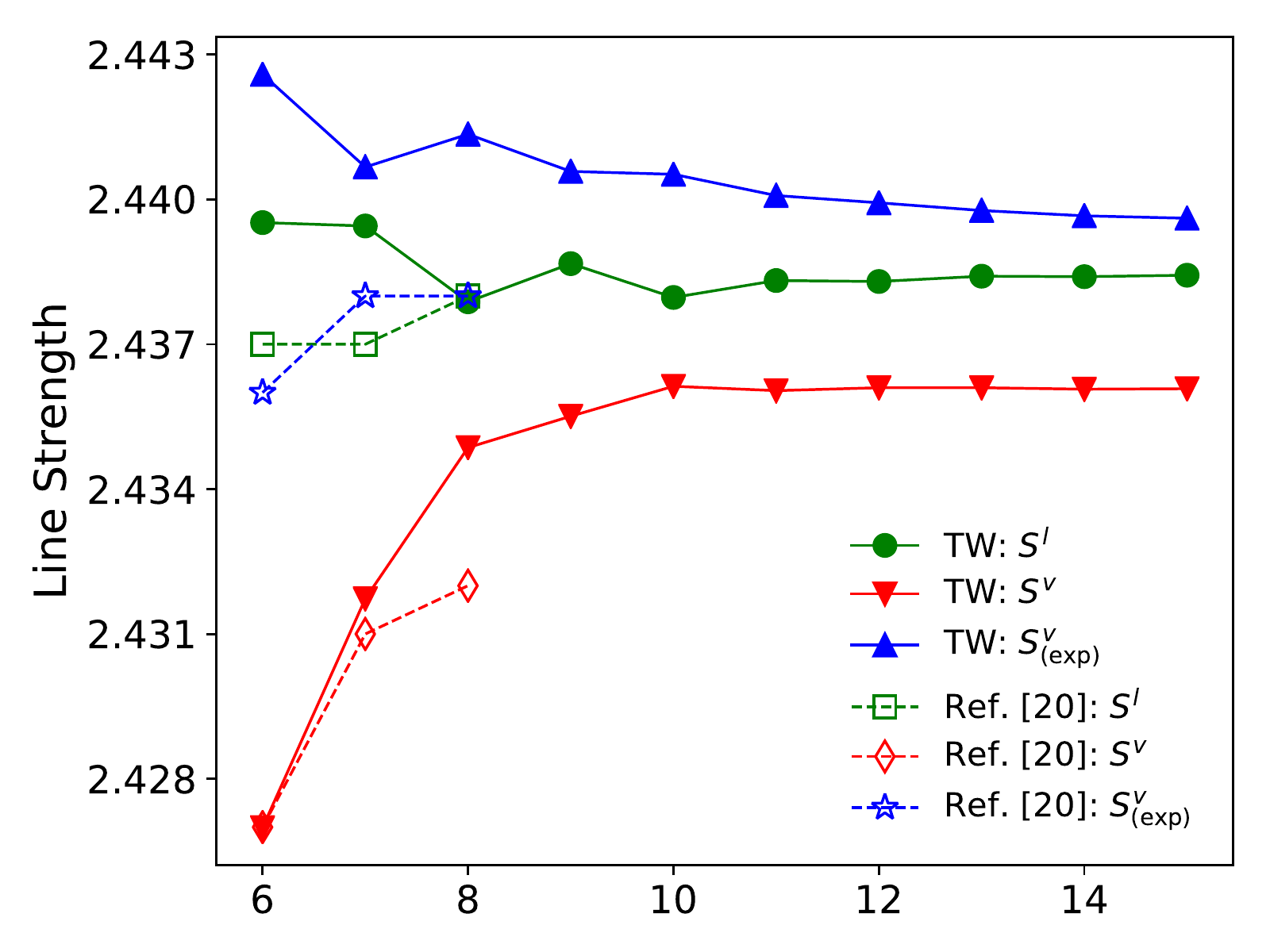}
\includegraphics[scale=0.52]{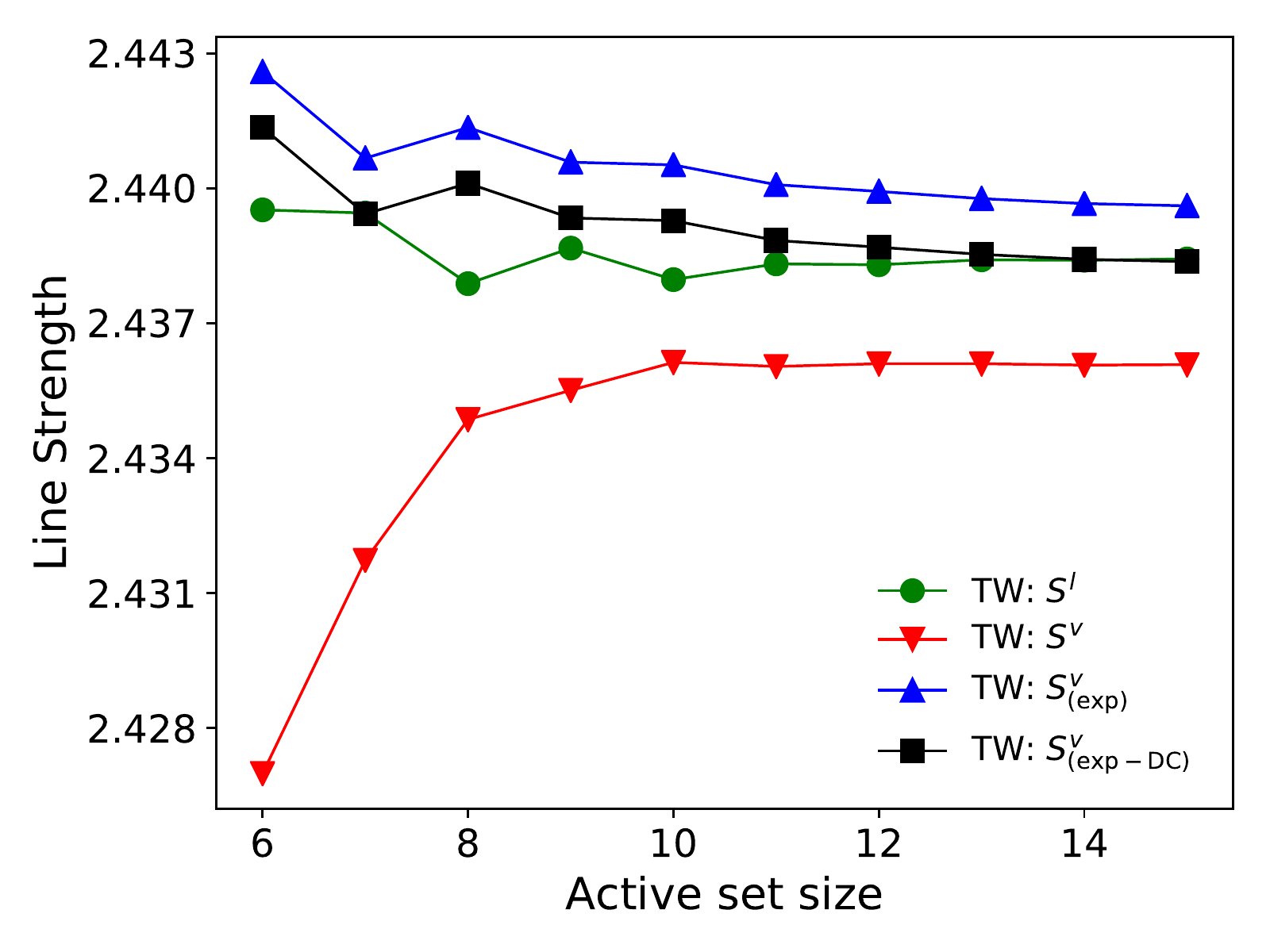}
\caption{Comparison of the line strengths evaluated according to the Model 1 (TW) as a function of
         active set size of the orbitals with the results of Ref.~\cite{Jonsson.57.4967.1998}. The green
         circles, red down triangles, and blue upper triangles display the present calculations
         in the length gauge $S^l$ as well as in velocity gauges before $S^v$ and after adjustment
         $S^v_{(\rm exp)}$ to the experimental transition energy $\omega_{\rm exp}$, respectively.
         The green hollow squares, red hollow diamonds, and blue hollow stars are corresponding
	     values taken from Ref.~\cite{Jonsson.57.4967.1998}.}
\label{pre-jon}
\end{figure}

We now explain how valence-valence~(VV), core-valence~(CV) and core-core~(CC) correlations were 
incorporated. We started to expand the basis set by adding the CSFs that are generated from the 
configurations $1s^2 nl n'l'$ which result from single and double (SD) excitations from outer 
shells of the reference configuration. These CSFs account VV correlations and the calculations are
named as VV calculations. 
To each layer of VV correlation calculations, we then added CSFs of the configurations 
$1s2s nl n'l' + 1s2p nl n'l'$ which arise from the single excitation from the $1s^2$ core with or
without another excitation from the valence shells. These added CSFs account for the CV
correlations and the calculations are called VV+CV. Now with each layer of VV+CV correlation
calculations, the correlations of two-electron excitations from the $1s^2$ core were included to account
for the CC correlations. These additional CSFs arise from the configurations 
$2s^2 nl n'l' + 2p^2 nl n'l'$ for the $^1S_0$  state and $ 2s2p nl n'l' $ for the $^1P_1$ state. 
These correlation calculations are named as VV+CV+CC calculations.
In all VV, VV+CV and VV+CV+CC calculations, the active set of orbitals was spanned by the orbitals with 
principal quantum number $n,n' \le 15 $ and with azimuthal quantum number $ l,l' \le 7 $.
Finally, the basis set of CSFs was expanded by appending CSFs with configurations arising from 
single, double, triple and quadruple (SDTQ) excitations from the reference configurations. In the SDTQ 
excitations
the number of CSFs increased very rapidly with the increasing number of orbitals in the active set 
which challenges the numerical stability and available hardware
resources. So the SDTQ excitations 
were limited only with $n,n' \le 5 $ and $ l,l' \le 4 $. These calculation were then extended 
with SD excitations with remaining layer of correlation orbitals with $n,n' \le 15 $ and $ l,l' \le 7 $.
We name this final set of calculations as VV+CV+CC:SDTQ. 

\subsection{Model 1}\label{model1}

In this model the VV and VV+CV calculations were performed by utilizing the OL scheme
for the ground and excited state, i.e., for both states orbitals are separately optimized, whereas for
the VV+CV+CC as well as VV+CV+CC:SDTQ calculations the spectroscopic orbitals and the correlation orbitals
with $ n = 3$ were simultaneously optimized using the EOL scheme. Then the calculations after $ n = 3$  
were extended with a separate set of correlation orbitals (the OL scheme). This model accounts for the 
correlations in a similar manner as those presented by J\"onsson and 
Froese Fischer \cite{Jonsson.57.4967.1998}.
The only difference is that J\"onsson and Froese Fischer performed VV+CV+CC:SDTQ calculations with SDTQ 
excitation until $ n = 3$ only and extended their calculations from $ n > 3 $ with VV+CV type of correlations only.

\begin{table*} 
\caption{Transition energies $\omega$ (cm$^{-1}$) and line strengths $S$ (a.u.) for the $1s^2 2s 2p$ $^1P_1 \,-\ 1s^2 2s^2$
$^1S_0$ transition in Be-like carbon as functions of the active set calculated within Model 1. The line strengths
in the length gauge ($S^l$) are compared with those in the velocity gauge ab initially calculated ($S^v$) and
after adjustment to the experimental energy $\omega_{\rm exp}$ ($S^v_{({\rm exp})}$) and to the
experimental-Dirac-Coulomb energy $\omega_{\rm exp-DC}$ ($S^v_{({\rm exp-DC)}}$). The experimental transition
energy is taken from Ref.~\cite{NIST}, while the experimental-Dirac-Coulomb energy is evaluated
by subtracting the Breit, recoil, and QED corrections from the experimental transition energy.}
\label{M1-corr} 
\begin{ruledtabular}
\begin{tabular}{lrcccc}
Active set                   &          $\omega$    &        $S^l$ &        $S^v$   & $S^v_{({\rm exp})}$ & $S^v_{({\rm exp-DC})}$ \\
\colrule                                                                                     
DHF                          &          112\,958    & 2.34092  &   1.65645     &     2.01753 &   2.01651    \\                               
$3s3p3d$                     &          104\,094    & 2.51432  &   2.36757     &     2.44884 &   2.44759    \\                               
$4s4p4d4f$                   &          103\,116    & 2.45884  &   2.38001     &     2.41568 &   2.41446    \\                               
$5s5p5d5f5g$                 &          102\,804    & 2.44978  &   2.40435     &     2.42565 &   2.42442    \\                               
$6s6p6d6f6g6h$               &          102\,680    & 2.43952  &   2.42699     &     2.44259 &   2.44135    \\                               
$7s7p7d7f7g7h7i$             &          102\,540    & 2.43945  &   2.43173     &     2.44067 &   2.43943    \\                               
$8s8p8d8f8g8h8i8k$           &          102\,488    & 2.43788  &   2.43486     &     2.44135 &   2.44011    \\                               
$9s9p9d9f9g9h9i9k$           &          102\,459    & 2.43867  &   2.43551     &     2.44058 &   2.43934    \\                               
$10s10p10d10f10g10h10i10k$   &          102\,444    & 2.43797  &   2.43613     &     2.44052 &   2.43928    \\                               
$11s11p11d11f11g11h11i10k$   &          102\,437    & 2.43832  &   2.43604     &     2.44008 &   2.43884    \\                               
$12s12p12d12f12g12h12i10k$   &          102\,432    & 2.43830  &   2.43610     &     2.43993 &   2.43869    \\                               
$13s13p13d13f13g13h13i10k$   &          102\,429    & 2.43841  &   2.43610     &     2.43977 &   2.43853    \\                               
$14s14p14d14f14g14h14i10k$   &          102\,427    & 2.43840  &   2.43607     &     2.43966 &   2.43842    \\                               
$15s15p15d15f15g15h15i10k$   &          102\,426    & 2.43843  &   2.43608     &     2.43961 &   2.43837    \\                               
Due to other models         &         -15          &          &               &             &              \\                               
Breit                        &         -3           &          &               &             &              \\                             
Recoil                       &         -13          &          &               &             &              \\                             
QED                          &         -10          &          &               &             &              \\                             
Total                        &         102\,385     &          &               &             &              \\                             
Exp                          &         102\,352     &          &               &             &              \\                             
Exp-DC                       &         102\,378     &          &               &             &              \\                             
\end{tabular}
\end{ruledtabular}
\end{table*}

In Fig. \ref{pre-jon} (upper plot) we compare the present calculations of Model 1 with J\"onsson and
Froese Fischer results \cite{Jonsson.57.4967.1998}. The line strength in length and velocity form
is plotted against increasing $n$ of the active set size defining the wave function expansion in
respective calculations. It is clearly evident that their two gauges agree perfectly at $n = 8$ once
the experimental energy adjustment was applied. Explicitly, the line strength calculated in
Ref.~\cite{Jonsson.57.4967.1998} amounts to 2.4376 in the length gauge and 2.4366 in the velocity
gauge after adjustment, which leads to a tabulated final value 2.4376(13). Despite we can not
explicitly reproduce these calculations, our evaluations show the similar effect at $n = 7$ active set
layer, where the results obtained in the length and velocity (adjusted) gauges approach each other.
However, when the active set size is further extended, one can clearly see that after $n = 7$ layer the
results first drift apart and then, again, approach each other but at some different position. These
observations lead us to the following conclusions. First, the agreement between gauges might be of
an accidental character and, therefore, second, the difference between results in the length and
velocity (adjusted) gauges should be very carefully used as a criterion for the error estimation.

Let us mention here another important observation. From the basic theory 
\cite{Savukov.62.052506.2000,Indelicato.69.062506.2004}, it is clear that the adjustment should 
be made to the transition energy which corresponds to the difference of the eigenvalues of the 
Dirac-Coulomb Hamiltonian in Eq.~(\ref{eq:H_DC}). Therefore, we calculate a so-called 
experimental-Dirac-Coulomb transition energy $\omega_{\rm exp-DC}$ as a difference of the experimental
value $\omega_{\rm exp}$ and the contributions beyond the Dirac-Coulomb approximation, i.e., the Breit 
interaction, recoil, and QED corrections. The $\omega_{\rm exp-DC}$ energy is, thus, experimentally 
deduced fully correlated Dirac-Coulomb transition energy. We compare in Table~\ref{M1-corr}
as well as in Fig.~\ref{pre-jon} (lower plot) the line strengths in the length gauge ($S^l$) and in the
velocity gauge (\textit{ab initio}) ($S^v$) with the adjusted to $\omega_{\rm exp}$ ($S^v_{({\rm exp})}$) and 
to $\omega_{\rm exp-DC}$ ($S^v_{({\rm exp-DC})}$) values. As one can see from this comparison the values 
adjusted to the experimental-Dirac-Coulomb energy are much close to the results in the length gauge.
In particular, for $n = 15$ layer the relative difference between the gauges amounts to $2 \times 10^{-5}$. 
However, as we mention at the end of Sec.~\ref{MCDHF} the agreement between the gauges cannot be uniquely 
used for the accuracy assessment.
For this reason, in the next subsections, we investigate also other correlation models.
\subsection{Model 2}\label{model2}
\begin{figure}[bh]
\includegraphics[scale=0.52]{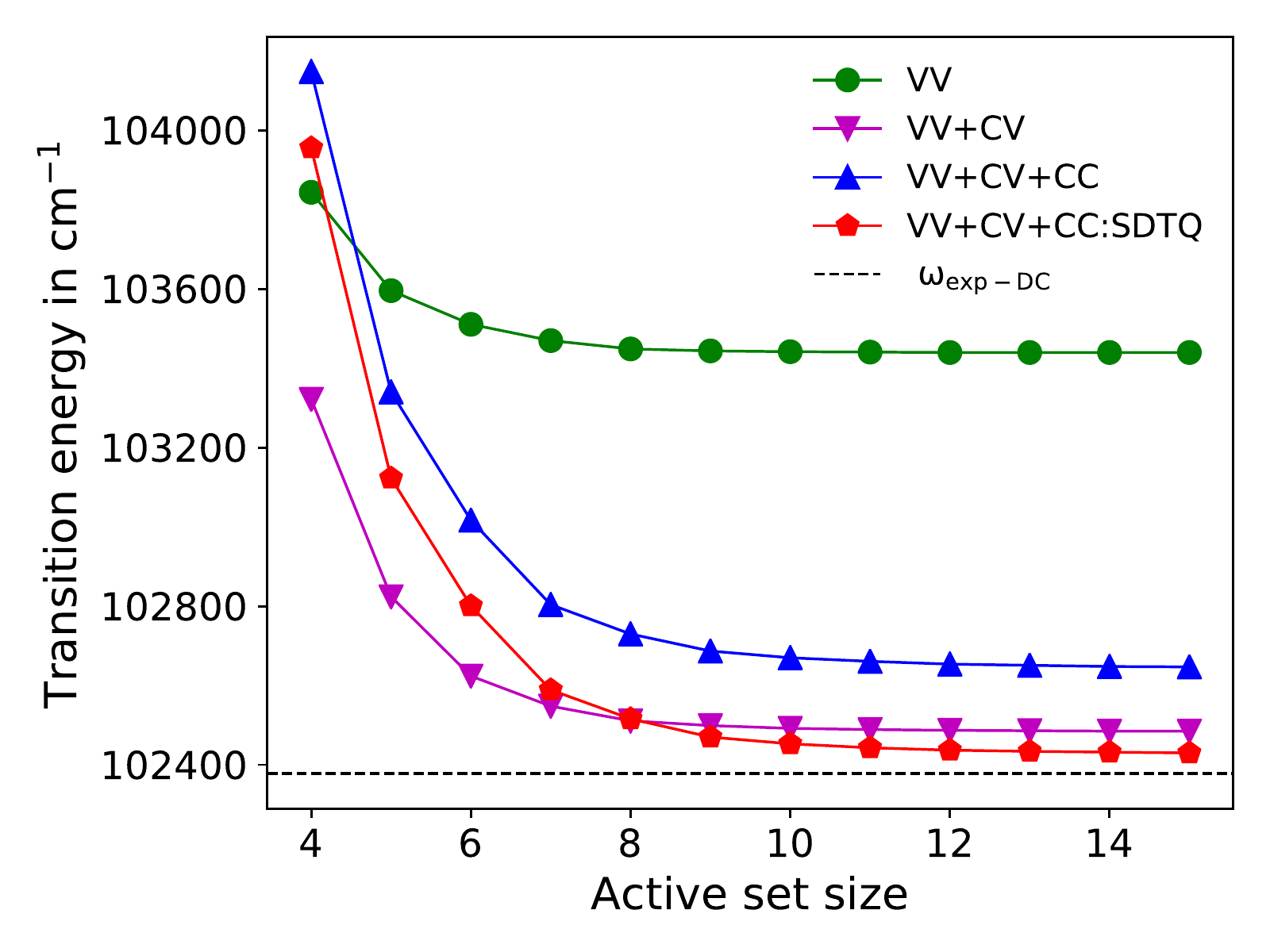}
 \caption{Extent of different correlations effects to the transition energy with respect to the
	  increasing $n$ of the active set size defining the wave function expansion for the Model
	  2 calculations. The green circles represent VV correlations, magenta down triangles
          represent VV+CV, blue upper triangles represent VV+CV+CC, red-pentagons represent
          VV+CV+CC:SDTQ correlations. Please see the text for the details of different type of
          the correlations.}
\label{diff.corr}
\end{figure}

Within this model, both the spectroscopic and the correlations orbitals were separately optimized using 
the OL scheme for all type of correlations. Hence generated orbitals for both states were not
quite orthogonal with each other, which makes the implementation of standard
Racah algebra difficult for the calculation of transition amplitude. To deal with this complication, a
transformation to a biorthonormal basis was applied together with the counter transformation of the 
expansion coefficients $c_{j}$ and $c'_{k}$ \cite{Olsen.52.4499.1995}.

The calculations within this model show the importance of a common set of orbitals
for the core correlations in the framework of MCDHF approach. For the CC effects, it is commonly accepted that these are more balanced if a common orbital basis is used for describing both the states involved in
the transition and hence resulting transition energies are more accurate, for details please see the
Ref.~\cite{Bieron.97.062505.2018,Filippin.94.062508.2016}. This is also obvious from Fig.~\ref{diff.corr} 
where the evaluation of different correlations
effects to the transition energy is shown with respect to the increasing $n$ of the active set size defining 
the wave function expansion for the Model 2 calculations. Here the blue upper triangles representing VV+CV+CC 
correlation results are worse than the magenta down triangles representing VV+CV correlations. 
However, it is obvious from the red-pentagons
representing VV+CV+CC:SDTQ in Fig.~\ref{diff.corr} that we get the best agreement of
the transition energy with the experiment when the TQ excitations are included with SD excitations 
in the CC correlations. The difference of the final values of 
the energy of VV+CV+CC:SDTQ calculations with Model 1 and Model 2 is only 0.004\%, whereas
the length form of the line strength from both models varies only by 0.01\%. 
The fact that TQ contributions are very important is also noticed from the results of Chen, Cheng and
Johnson~\cite{Chen.64.042507.2001} who have also used TQ excitations in building a common set of 
orbitals in their RCI calculations based on B-spline basis. 

\subsection{Model 3}\label{model3}

Within this model, both the spectroscopic orbitals and correlations orbitals were simultaneously optimized for 
the ground and excited states using the EOL scheme for all type of correlations. The so obtained orbitals
for both states were orthogonal to each other. 
As it has been highlighted by Chen, Cheng and Johnson~\cite{Chen.64.042507.2001} and 
Savukov~\cite{Savukov.70.042502.2004} that small orbital overlap
corrections due to nondiagonal set of orbitals for the initial and final state should not be ignored.
Our correlation Model 3 helped to address this issue.  

\subsection{Model MR}\label{modelMR}

In this model, the set of spectroscopic reference configurations were expanded to account 
the missing correlations due to limited SDTQ excitations. We name it as
multi-reference~(MR) model. The configurations in MR are expanded in such a way that the CSFs for 
the MR set had the largest expansion coefficients 
in the wave functions that were generated by VV+CV+CC:SDTQ calculations of Model 3. 
For the $ ^1S_{0}$ ground state the resulting MR set was 
\{$ 1s^{2}2s^{2}, 1s^{2} 2p^{2}, 1s^{2} 3p^{2},2s^{2} 3s^{2}, 2s^{2} 3p^{2}, 1s^{2} 3d^{2}$\} and 
for the $^1P_{1} $ excited state the resulting MR set was
\{$ 1s^{2} 2s 2p, 1s^{2} 2p3d, 2s2p3s^2, 2s2p3p^2 $\}. All the 
orbitals occupied in MR set were spectroscopically treated in the lowest order of approximation. 
Then the correlation orbitals were treated in the same way as those of Model 3 using the EOL scheme.

\subsection{Models: summary}

Our approach with either a common or two separate set of orbitals for the ground and excited states
combines the strengths and weaknesses of the previous calculations which provide
the uncertainty of the order of $10^{-3}$
\cite{Jonsson.57.4967.1998,Chen.64.042507.2001,Savukov.70.042502.2004}. 
The orbitals in the common set for both states are orthogonal to each other and there is 
no orbital overlap for the evaluation of the transition amplitude. At the same time, our procedure 
of two different sets of orbitals for each state has the advantage that the electron relaxation 
effects are automatically included to a large extent. 
\begin{figure}[bh]
\includegraphics[scale=0.52]{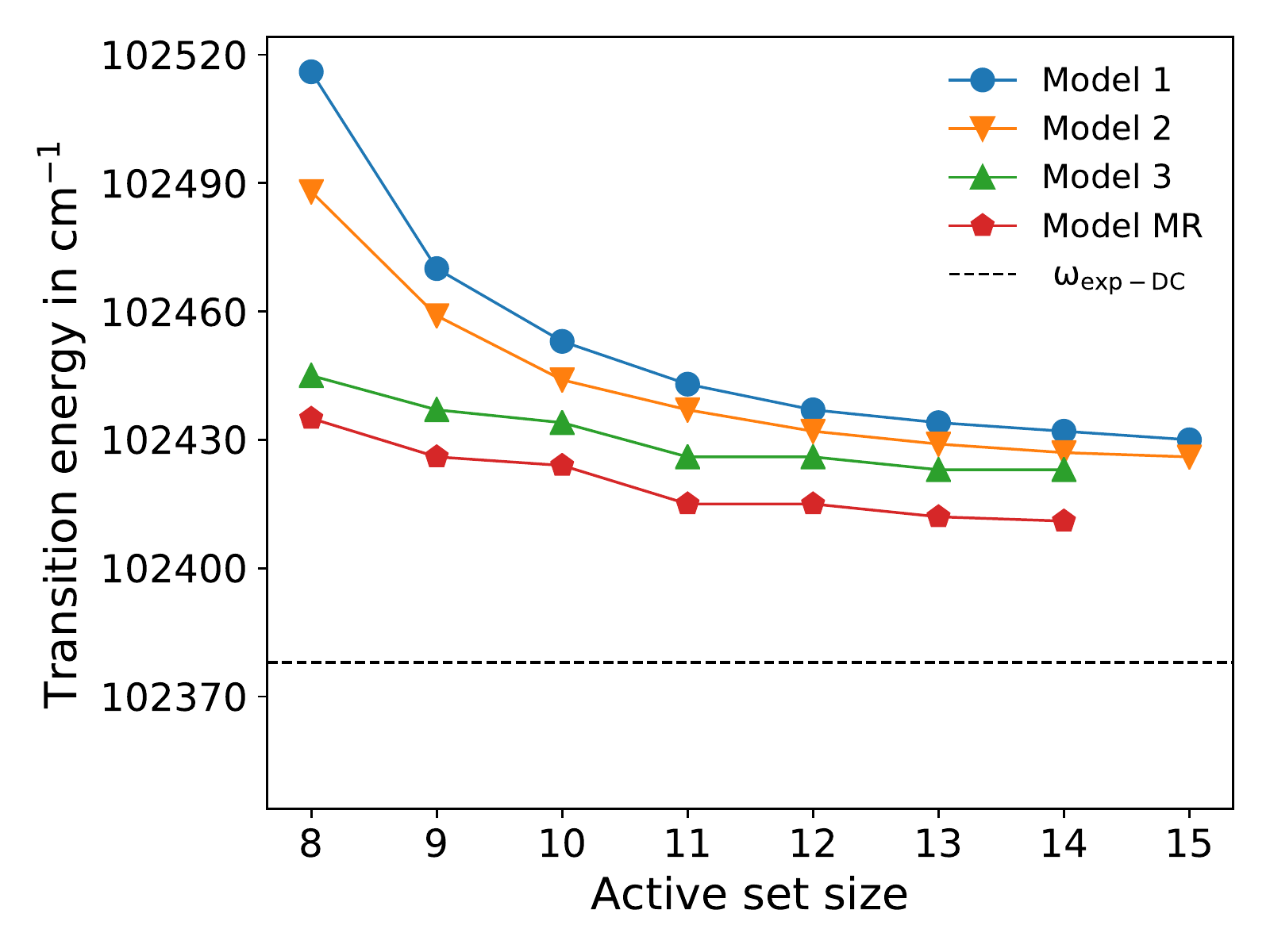}
 \caption{Convergence of transition energy with respect to the increasing $n$ of the active set size 
	  defining the wave function expansion for the VV+CV+CC:SDTQ calculations of different 
	  correlation models. Please see the text for the details of the 
	  VV+CV+CC:SDTQ calculations.}\label{diff.Es}
\end{figure}
In all our correlation models the overall convergence trends and behavior of the inner 
and outer electron correlations are consistent with each other. 
In Fig.~\ref{diff.Es} we present the convergence of the transition energy with 
respect to the increasing $n$ of the active set size defining the wave function expansion 
for the VV+CV+CC:SDTQ calculations from all the correlation models under present study. For the 
Model 3 and MR we could not get the converged orbitals for $n = 15$. The first three models vary just with a 
difference of maximum of 6 cm$^{-1}$. But for Model MR we get 15 cm$^{-1}$ better results and this
is obviously due to the inclusion of higher-order correlations in this model.

\begin{figure}[p]
\subfloat{\includegraphics[scale=0.52]{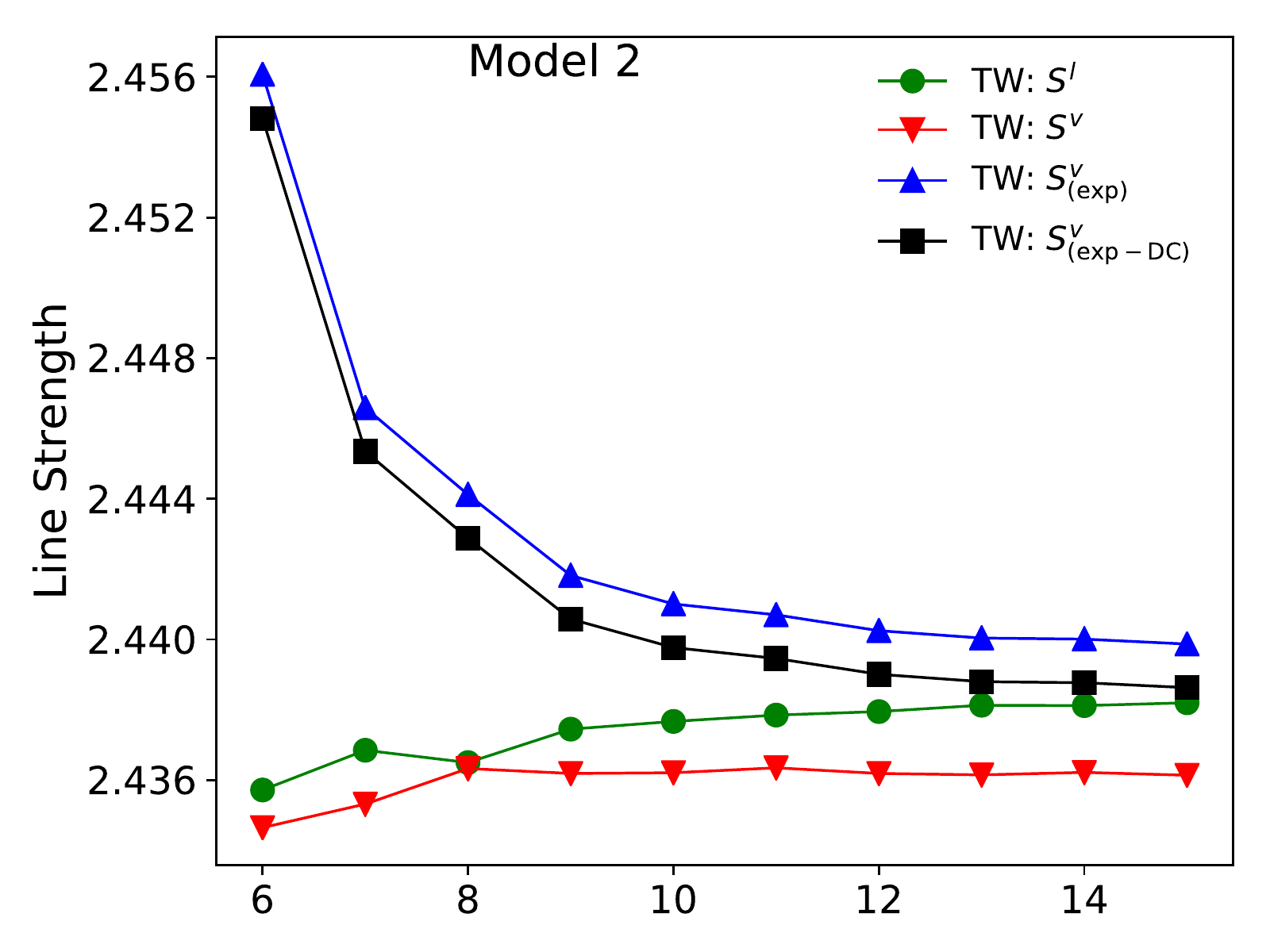}}\\ 
\subfloat{\includegraphics[scale=0.52]{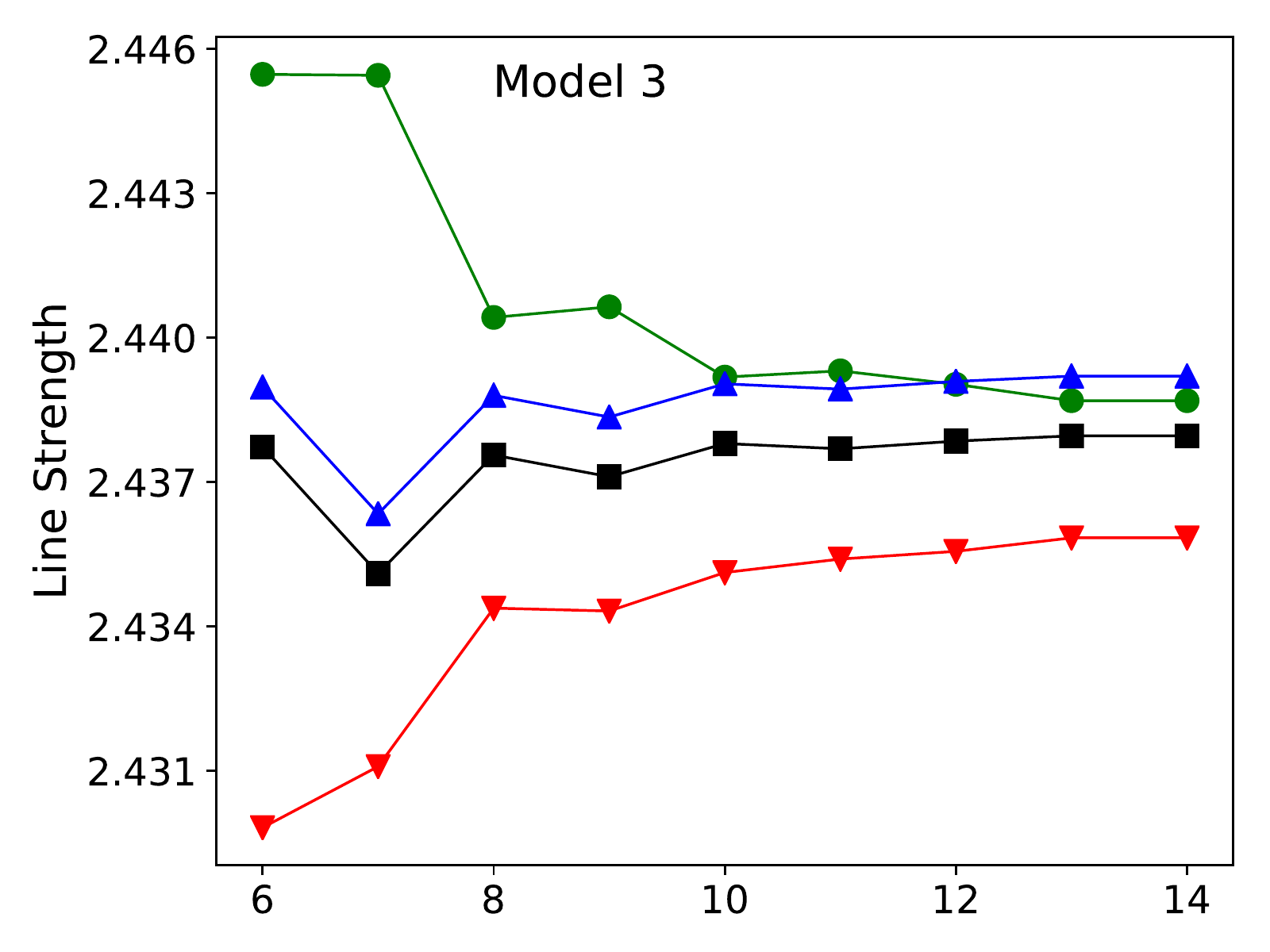}}\\
\subfloat{\includegraphics[scale=0.52]{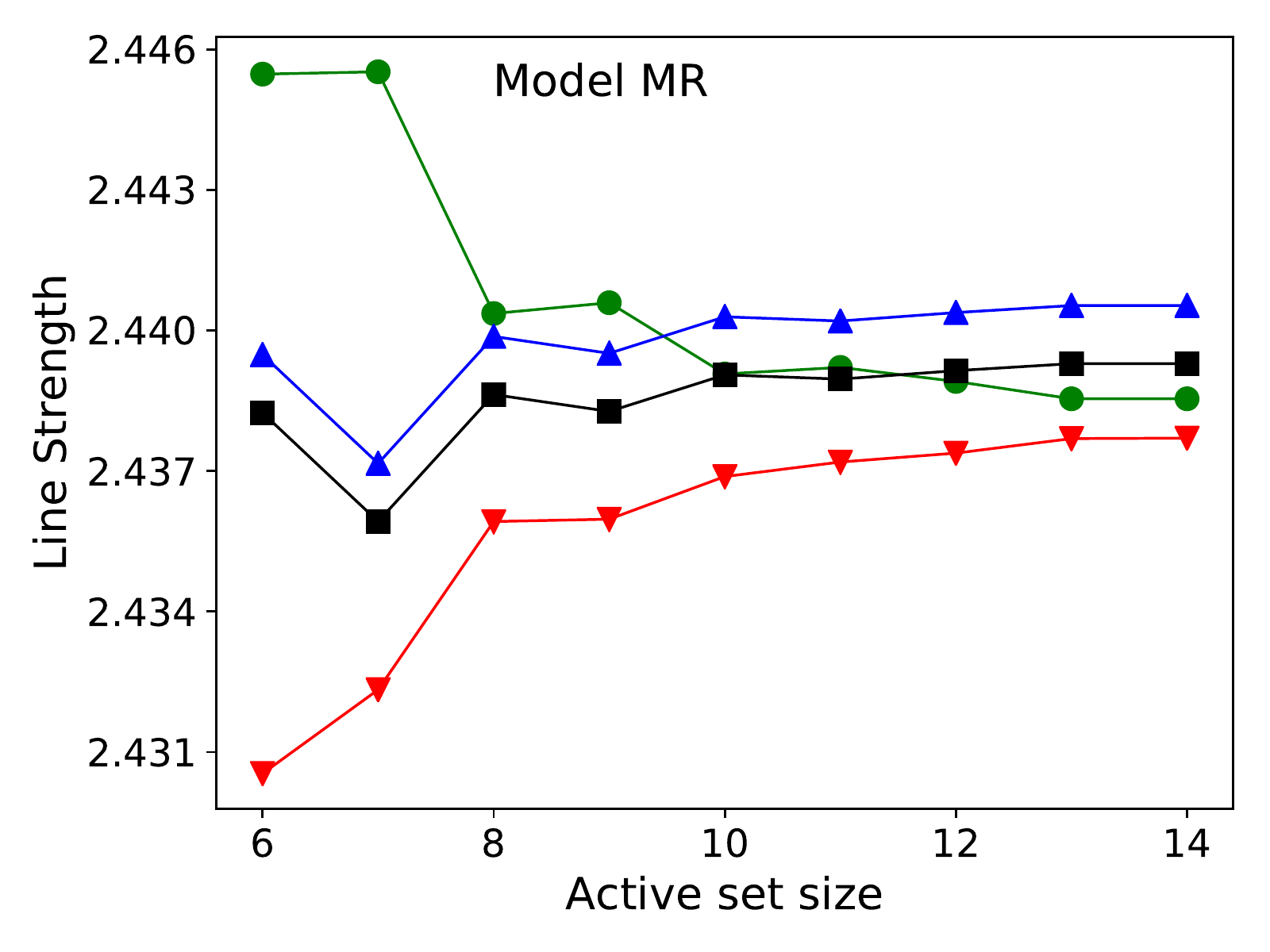}}
\caption{Line strength for the for the VV+CV+CC:SDTQ calculations from the Model 2, Model 3 and Model MR
 plotted similarly as that of Fig.~\ref{pre-jon}. In each sub-figure, the green circles
	 are the line strength $S^l$ in length form, red down triangles are \textit{ab initio} $S^v$ velocity from,
	 blue upper triangles are $S^v_{(\rm exp)}$ velocity form adjusted the experimental transition energy
	 $\omega_{\rm exp}$ and black squares are $S^v_{({\rm exp-DC})}$ velocity from adjusted the 
	 experimental-Dirac-Coulomb transition energy $\omega_{\rm exp-DC}$ (see text).}\label{linestrengths}
\end{figure}

\begin{figure}[h]
\includegraphics[scale=0.52]{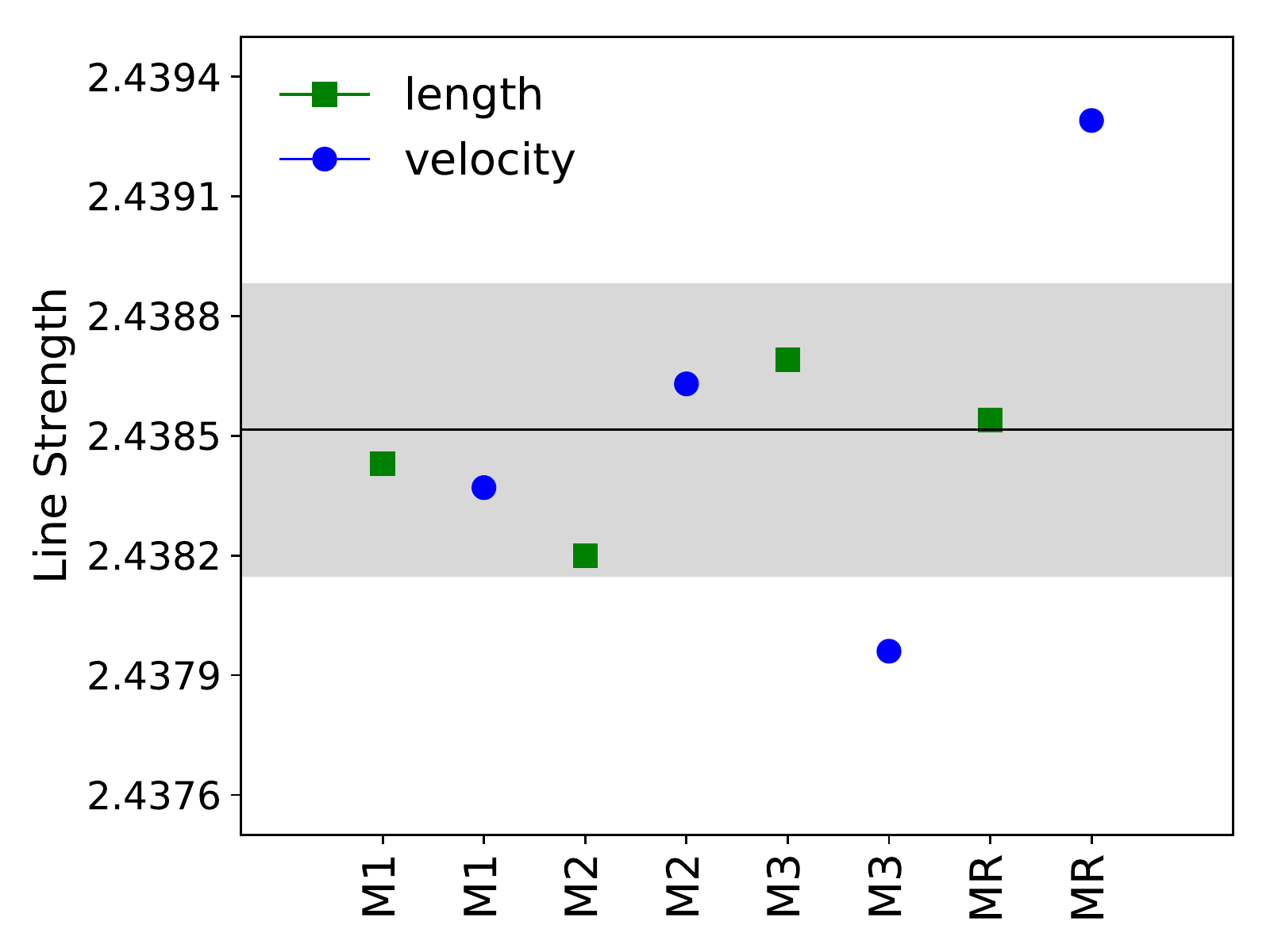}
\caption{Line strength of the $1s^2 2s 2p$ $^1P_1 \,-\ 1s^2 2s^2$
$^1S_0$ transition in Be-like carbon plotted against the present Models.
     Green squares correspond to the length-form values and blue circles to the velocity-form values.
	 The black solid line is the average of the both length 
	 and velocity forms. The gray shaded region is one standard deviation with respect to the black line.}
	 \label{S-diff-mod}
\end{figure}

In Fig.~\ref{linestrengths} we present the line strength for the VV+CV+CC:SDTQ correlations calculated
within Model 2, 3, and MR in a similar way as explained in Sec.~\ref{model1} and Fig.~\ref{pre-jon}
(lower plot). From these plots, one can clearly see that in all models the line strength in the velocity
adjusted to the experimental-Dirac-Coulomb energy agrees with the length gauge result much better than
the adjusted to the pure experimental energy. This also confirms our expectations originated from the
basic principles as stated at the end of Sec.~\ref{model1}. In order to get the final (Dirac-Coulomb)
line strength value the results of the length gauge and adjusted to the experimental-Dirac-Coulomb
energy velocity gauge obtained at the maximum active set size are analysed. The employed data from all
the correlation models are summarized in Table~\ref{table:modelsS} and Fig. \ref{S-diff-mod}. As one can see
from these data, despite the extraordinary agreement between gauges, e.g., in Model 1, one can not use it
for the uncertainty estimation. The reason for this has been explained at the end of Sec.~\ref{MCDHF}
and confirmed now by the calculations in other models, which predict quite larger spread of the results
than given by the difference of the gauges. We take an average of these scatter of the line strength data to predict the final
value of the line strength and take one standard deviation of these scatter of the line strength data to predict the uncertainty in the results. As a result,
present line strength accounting only the correlations is 2.43851(37). This is represented as a black solid line in Fig.~\ref{S-diff-mod}, whereas in this figure the uncertainty is shown as a gray shaded region.
We find rather conservative to assess the uncertainty as one standard deviation of the scattered data.
Such kind of an error estimation is further supported
by the fact that it covers all the values obtained in the length gauge, which is known to be more
reliable.

\begin{table}[h] 
\caption{$\omega$ is the Dirac-Coulomb transition energy (cm$^{-1}$). $S^l$ and $S^v_{({\rm exp-DC})}$ are
         the Dirac-Coulomb line strengths in the length gauge and in the velocity gauge adjusted to
         $\omega_{\rm exp-DC}$ energy (see text), in a.u.}
\label{table:modelsS}
\begin{ruledtabular}
\begin{tabular}{lccc}
Label    &         $\omega $    & $S^l$&  $S^v_{({\rm exp-DC})}$   \\
\colrule
Model 1   & 102426    &      2.43843    & 2.43837      \\
Model 2   & 102430    &      2.43820    & 2.43863      \\
Model 3   & 102423    &      2.43869    & 2.43796      \\
Model MR  & 102411    &      2.43854    & 2.43929      \\
Final     &           &\multicolumn{2}{c}{2.43851(37)} \\
\end{tabular}
\end{ruledtabular}
 \end{table}

\section{nuclear recoil correction}\label{rec}

Once the line strength is calculated, including all the
major correlation contributions, the finite nuclear mass 
(nuclear recoil) contribution was added as a correction given as 
\begin{equation}\label{delta S_rec}
\Delta S_{\mathrm {rec}} = \Delta S_{\mathrm {rec,en}} + \Delta S_{\mathrm {rec,wf}}  + 
            \Delta S_{\mathrm {rec,op}},   
\end{equation}
where the first two terms on the right side in Eq.~(\ref{delta S_rec}) are the 
corrections to the line strength due to nuclear-recoil contributions to the energy and
wave functions. These corrections were calculated using the relativistic configuration interaction~(RCI) program of
\textsc{Grasp2K} \cite{Jonsson.184.2197.2013}. Here the lowest order nuclear motional corrections, 
namely the normal mass shift (NMS) term based on Dirac kinetic energy operator  
\begin{equation}\nonumber
\hat{H}_{\mathrm {NMS}}=\frac{1}{M}\sum_{i=1}^{N}\Bigl[c\boldsymbol{\alpha}_{i}\cdot 
						     \boldsymbol{p}_{i}+(\beta_{i}-1)c^{2}\Bigr],
\end{equation}
where the $M$ is the nuclear mass, and mass polarization term named as the specific mass shift (SMS) 
\begin{equation}\nonumber
\hat{H}_{\mathrm {SMS}}=\frac{1}{M}\sum_{i<j=1}^{N} \bm{p}_{i} \cdot \bm{p}_{j}, 
\end{equation}
were added to the Dirac-Coulomb Hamiltonian, Eq.~(\ref{eq:H_DC}). The additional relativistic corrections to the recoil operators are of the order $(Z\alpha)^2 $ \cite{Karshenboim.56.4311.1997}. For $Z = 6$, these corrections are in the order of $10^{-6}$, which is far below than present level of uncertainty. However, the relativistic corrections must be taken into considerations for the future studies of higher $Z$.

It is also important, however, to take into account the recoil correction to the transition operator,
i.e., the third term in Eq.~(\ref{delta S_rec}). Previously, it was considered in
Refs.~\cite{Fried.29.574.1963,Yan.52.R4316.1995,Karshenboim.56.4311.1997,Yan.57.1652.1998} for 
the $E1$ transitions and in Ref.~\cite{Volotka.48.167.2008} for the $M1$-decay. 
Starting from the nonrelativistic Hamiltonian for $N$ electrons and the nucleus,
we obtain the following recoil corrections to the $E1$ transition operator:
\begin{equation}
\label{eq:Tl_rec}
\Delta\boldsymbol{T}^l_{\rm rec} = -\frac{Z-N}{M} \sum_{i=1}^{N} \bm{r}_{i}
\end{equation}
in the length gauge and
\begin{equation}
\label{eq:Tv_rec}
\Delta\boldsymbol{T}^v_{\rm rec} = -\frac{Z}{M}\frac{1}{\omega} \sum_{i=1}^{N} \bm{p}_{i}
\end{equation}
in the velocity gauge. From these expressions one can easily come to the corresponding
corrections
to the line strength
\begin{eqnarray}
\Delta S^l_{\rm rec,op} &=& 2 {\rm Re} \left\{\langle\Psi(\Gamma; \pi J) 
 ||\boldsymbol{T}^l||\Psi(\Gamma'; \pi' J')\rangle \right.\nonumber\\
 &\times&\left. \langle\Psi(\Gamma; \pi J)||\Delta\boldsymbol{T}^l_{\rm rec}||\Psi(\Gamma'; \pi' J')
                                       \rangle\right\}\nonumber\\
 &\approx& 2 \frac{Z-N}{M}S^l
\end{eqnarray}
and
\begin{eqnarray}
\Delta S^v_{\rm rec,op} &=& 2 {\rm Re} \left\{
 \langle\Psi(\Gamma; \pi J)||\boldsymbol{T}^v||\Psi(\Gamma'; \pi' J')\rangle\right.\nonumber\\
 &\times&\left.\langle\Psi(\Gamma; \pi J)||\Delta\boldsymbol{T}^v_{\rm rec}||\Psi(\Gamma'; \pi' J')
                                       \rangle\right\}\nonumber\\
 &\approx& 2 \frac{Z}{M}S^v\,.
\end{eqnarray}
\begin{table}[h]
\caption{The recoil corrections to the line strength originated from the energy and wave functions change,
         $\Delta S_{\rm rec,en+wf}$, as well as due to the transition operator $\Delta S_{\rm rec,op}$
         calculated in the length and velocity gauges. The total gauge invariant recoil correction is
         presented in the last line. The values are in a.u.}\label{table:rec}
\begin{ruledtabular}
\begin{tabular}{lrr}
Rec. correction            & length  &   velocity \\
\colrule
$\Delta S_{\rm rec,en+wf}$ & 0.00000 &$-$0.00089  \\
$\Delta S_{\rm rec,op}$    & 0.00045 &   0.00134  \\ 
Total                      & 0.00045 &   0.00045  \\ 
\end{tabular}
\end{ruledtabular}
\end{table}
In Table~\ref{table:rec} different recoil contributions due to the energy, wave functions, and operator are
presented in the length and velocity gauges. Only with the term due to the change of the operator
included the total recoil correction is gauge invariant. In view of this, we would recommend to introduce
this contribution also to the next GRASP update.


%
\section{discussion and conclusion}\label{diss}
\begin{table}[bh] 
\caption{Comparisons between different calculations and experiments for the line strength of the
         $1s^2 2s 2p$ $^1P_1 \,-\ 1s^2 2s^2$ $^1S_0$ transition and the lifetime of the
         $1s^2 2s 2p$ $^1P_1$ excited state in Be-like carbon.}\label{table:Be_S_E1_comp}
\begin{ruledtabular}
\begin{tabular}{llc}
 $S$ [a.u.] & $\tau$ [ns] & Ref. \\
\colrule
\multicolumn{3}{c}{Theories}\\
   2.434(6)    &  0.5673(13)   &\cite{Fischer.49.323.1994}        \\               
   2.435(6)    &  0.5671(13)   &\cite{Ynnerman.51.2020.1995}      \\               
   2.4376(13)  &  0.56650(30)  &\cite{Jonsson.57.4967.1998}       \\               
   2.057       &  0.6713       &\cite{Safronova.59.286.1999}      \\               
   2.4377(10)  &  0.56648(23)  &\cite{Chen.64.042507.2001}        \\               
   2.4390(24)  &  0.56618(55)  &\cite{Savukov.70.042502.2004}     \\               
   2.436       &  0.5669       &\cite{Wang.234.40.2018}           \\               
   2.43926(40) &  0.56612(9) & Present work                     \\               
\colrule
\multicolumn{3}{c}{Experiments}   \\
   2.09(1)       & 0.66(3)       &\cite{Heroux.180.1.1969}                 \\        
   2.16(2)       & 0.64(6)       &\cite{Poulizac.4.191.1971}               \\        
   2.09(2)       & 0.66(7)       &\cite{Poulizac.8.40.1973}                \\        
   2.76(1)       & 0.50(3)       &\cite{Chang.211.300.1977}                \\        
   2.42(1)       & 0.57(2)       &\cite{Reistad.34.151.1986}          \\             
   2.426(45)     & 0.569(10)     &\cite{Reistad.34.2632.1986}       \\               
\end{tabular}
\end{ruledtabular}
 \end{table} 
\begin{figure}[bh]
\includegraphics[scale=0.5]{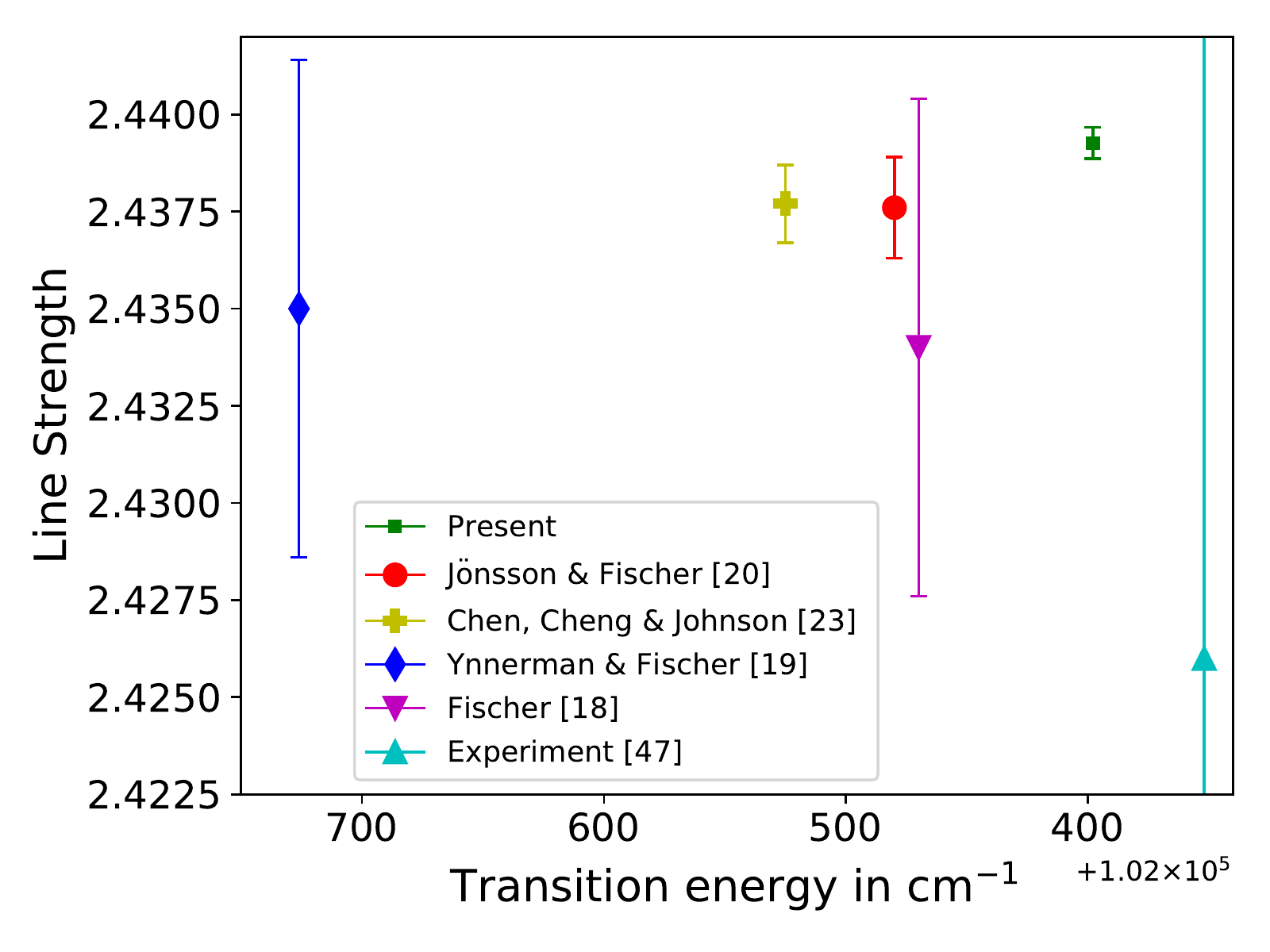}
\caption{A comparison of the present line strength of the 
	 $ 1s^{2} 2s 2p$  $^1P_{1} \,-\ 1s^{2} 2s^{2}$ $^1S_{0} $ transition in Be-like carbon 
	 with other theories and experiment.}
	 \label{comp}
\end{figure}

With the discussion above, we can obtain the final value of the line strength. In order to do so, we add to the
Dirac-Coulomb value 2.43851(37) from Sec.~\ref{models} the recoil correction
$\Delta S_{\rm rec} = 0.00045$ calculated in the
previous section. In addition, we have to consider also other effects, such
as the Breit interaction as well as QED. The Breit contribution has been calculated as follows. The
frequency-independent Breit Hamiltonian has been added to the Dirac-Coulomb Hamiltonian given by 
Eq.~(\ref{eq:H_DC}).
Then the RCI calculations have been performed within the correlation Model 1. Comparing further the
obtained results with the corresponding Dirac-Coulomb values we get for the Breit contribution $\Delta
S_{\rm Breit}$ $0.00030$ and $-0.00001$ in the length and velocity gauge, respectively. Based on an analysis
of the Breit contribution in the intercombination transition $2s 2p$ $^3P_1 \,-\ 2s^2$ $^1S_0$ in Be-like
carbon Ref.~\cite{Chen.64.042507.2001} and on arguments presented in 
Refs.~\cite{Johnson.35.255.1995,Derevianko.58.4453.1998},
we attribute this difference to the negative-energy corrections.
That means that the gauge invariance of the Breit contribution should
be restored when the negative-energy states will be accurately taken
into account. On the other hand, it was demonstrated
\cite{Johnson.35.255.1995,Derevianko.58.4453.1998}, that the negative-continuum affects
dominantly only the result in the velocity gauge, while the result of the length gauge remains
stable. In view of this, we add the Breit correction calculated in the length gauge and with the
50\% uncertainty, $\Delta S_{\rm Breit} = 0.00030(15)$, to our final value.
The remaining QED correction is estimated as $\alpha
(\alpha Z)^2 {\rm ln}(\alpha Z)^{-1}$ \cite{Ivanov.210.313.1996,Sapirstein.69.022113.2004}
to be $4 \times 10^{-5}$, which is much smaller than our uncertainty. As a result, our final value for the line strength
reads 2.43926(40), where the uncertainty is coming from the correlations and Breit contribution.

Once the line strength is calculated, it is straightforward to get the weighted oscillator strength $ gf $:
%
\begin{equation}
gf = \frac{2}{3}\omega S,
\end{equation}
and the lifetime of the $1s^2 2s 2p$ $^1P_1$ excited state:
\begin{equation}
\tau = \frac{3g}{4}\frac{c^3}{\omega^3 S}\,,
\end{equation}   
where $g$ is the weight of the upper state. Here the conversion to the lifetime form the line strength and vice versa is performed by
using the experimental energy, $\omega = 102 352.04$ cm$^{-1}$ \cite{NIST}. We note that
the present uncertainty in the lifetime is only due to calculated line strength, since the
uncertainty of the transition energy is expected to be much better than 1 cm$^{-1}$ and
this is far below the uncertainty of the line strength.

Fig.~\ref{comp} and Table~\ref{table:Be_S_E1_comp} compare the present results of
calculated line strength and lifetime with other theories and experiments. Note that
in the respective papers the values of oscillator strength is provided. We have converted
the oscillator strength to the line strength using the energies mentioned in the respective
papers. In Fig.~\ref{comp}, the experimental line strength reported in Ref.~\cite{Reistad.34.2632.1986}
is plotted as a function of the experimental energy taken from the NIST database \cite{NIST}.
It is clear from Fig.~\ref{comp} that our calculated energy is the closest to the experimental
one. The present line strength or lifetime agrees very well with the CI+MBPT calculation
of Savukov \cite{Savukov.70.042502.2004} and is in a fair agreement with the large-scale MCDF
calculation of J\"onsson and Froese Fischer \cite{Jonsson.57.4967.1998} and the CI result of
Chen, Cheng and Johnson \cite{Chen.64.042507.2001}.

It is obvious from the Table~\ref{table:Be_S_E1_comp} that all the theoretical lifetimes except
of Ref.~\cite{Safronova.59.286.1999} are inside the error bar of the best available experimental
lifetime of 0.569(10) ns \cite{Reistad.34.2632.1986}. However, the uncertainty of this measurement
is still too large to distinguish between different theories. Therefore, we hope that the proposed 
experiment in Ref.~\cite{Rothhardt.2019} will provide a new benchmark for testing the
theories in the case of Be-like carbon.
  
%
%
In conclusion, we have presented high precision atomic calculations of the line strength of the
$1s^2 2s 2p$ $^1P_1 \,-\ 1s^2 2s^2$ $^1S_0$ spin allowed $E1$ transition in Be-like carbon. We have
utilized the state-of-the-art multiconfiguration Dirac-Hartree-Fock method. In these systematically
enlarged wave functions, we incorporated higher-order electron correlations, where the orbital
relaxation and overlaps are taken into account by using separate and simultaneous sets of
relativistic orbitals in the active set. This helped us to reliably estimate the uncertainty
of the obtained line strength. Moreover, the finite nuclear mass correction to the line strength
is calculated by correcting the energy, wave functions as well as the transition operator. The
achieved relative uncertainty of the line strength amounts to $10^{-4}$, which represents
a reliable theoretical benchmark of the $E1$ line strength in a view of upcoming high precision
lifetime measurement of the $ 1s^{2} 2s 2p$  $^1P_{1} $ state of Be-like carbon.

Extensions of current studies to heavier Be-like ions allow 
us to improve the theoretical accuracy of transition rates. The given (numerical) uncertainty together with 
the high precision experiments will allow an alternative spectroscopic test
than the energy alone and will provide further insight into the atomic structure of many-electron atoms and ions.
\end{document}